\documentstyle[12pt]{article}
\def\wa{\widetilde{A}}
\def\p{\partial}
\def\tr{\,{\rm tr}\,}
\def\am{A_{\mu}}

\def\an{A_{\nu}}

\def\al{A_{\lambda}}

\def\be{\begin{equation}}
\def\ee{\end{equation}}
\def\bea{\begin{eqnarray}}
\def\eea{\end{eqnarray}}

\def\e{\epsilon}
\def\a{\alpha}
\def\tf{\widetilde{\varphi}}
\def\d{\delta}
\title{ On the canonical quantization of anomalous SU(N) chiral Yang-Mills
models}

\vspace{.7cm}
\author{ \mbox{}
Cornelius Sochichiu \thanks{e-mail: sochi@class.mian.su}
\mbox{} \\ Steklov Mathematical Institute
\vspace{-0.1cm} \mbox{}  \\ Vavilov st.42, GSP-1, 117966 Moscow, Russia
\vspace{-0.1cm} \mbox{} \\  \centerline{and}
\vspace{-0.1cm} \mbox{} \\ Physical Dept. of Moscow State University \\ Russia
\date{}}

\begin{document}

\maketitle

\section{Introduction}
The chiral gauge models are known to suffer from anomalies leading to the
inconsistency of the quantum theory \cite{1,2,3,4,5,6,7}.
This inconsistency may be avoided by modification of the classical action
consisting in adding to it  the Wess-Zumino (WZ) action modeling the
anomaly \cite{4,5,6,8}.

If the symplectic form of the modified action is nondegenerate, we have
a well-defined action in which anomalous contribution of old fields is
canceled with the similar one of new WZ fields. As a result, one has a
unitary gauge theory  on the physical subspace.  However, in many
interesting cases the symplectic form is degenerate and one must deal with
Dirac machinery for quantization of constrained systems. It is
important to know whether this machinery preserves the gauge invariance of
the theory or not. Due to the fact that WZ action is the first order one
in its fields the symplectic form must be degenerate at least for all
odd-dimensional groups such as $SU(2k)$, $dim\, SU(2k)=4k^2-1$.  The
particular case of two dimensional $SU(2)$ and four dimensional
$SU(3)/SO(3)$ models was considered in \cite{fss2,fss3}. In this paper we
consider the general case of SU(N) group with degenerate symplectic form
in four space-time dimensions.

The plan of paper is as follows. In the next section we briefly review the
method proposed by Faddeev and Shatashvili for quantization of anomalous
Yang-Mills theory. In the third section, we describe its generalization
for degenerate WZ actions and show that this generalization leads to gauge
invariant quantum theory.

\section{Anomalous SU(N) Yang-Mills model}
Consider the four dimensional Yang-Mills model described by the following
classical action:
\be
S_{cl}=\int d^4x\,(-\frac {1}{4} (F_{\mu\nu}^{a})^{2} +
i\bar{\psi} \hat{\nabla}(A) \psi),
\label{1}
\ee
where $\hat{\nabla}(A) = \gamma^\mu (\p_\mu + A_{\mu}) $ and $\psi \equiv
\frac12 (1+\gamma_5)\psi$ is chiral fermion field in the fundamental
representation of the gauge group generated by $\lambda^a$ satisfying
\be
[\lambda^a,\lambda^b]=f^{abc}\lambda^c, \quad \frac12\tr
\lambda^a\lambda^b=-\d^{ab}.
\ee
Due to the gauge invariance the action (\ref{1}) have a set of first class
constraints $G^a({\bf x})$ with the following
Poisson bracket algebra
\be
\{G^a({\bf x}),G^b({\bf y})\}=f^{abc}G^c({\bf x}) \delta({\bf x-y})
\label{2}
\ee
\be
\{H,G^a({\bf x}) \}|_{G=0}=0
\label{3}
\ee
This enables one to impose a gauge fixing condition. Let us use the
temporal gauge ($A_0=0$). In this case one has constraints $G$ dropped out
from the classical action which now become a nondegenerate one. This
action can be quantized in a usual way. The loosed Lorentz invariance will
be restored on the physical subspace of the Hilbert space consisting of
vectors satisfying
\be
\hat{G} |ph> =0
\label{4}
\ee
Writing (\ref{4}) one tacitly suppose that the (quantum) algebra of
operators $G$ will be identical to (\ref{2}) and (\ref{3}) with substitution
$ \{\, ,\} \rightarrow i[\, ,]$. Unfortunately that is not the case for
anomalous theories. Quantum corrections destroy the gauge invariance in
such theories.

Indeed the fermionic determinant
\be
\det \hat{\nabla}(A)= \int \hbox{e}^{iS_{cl} (A,\bar{\psi},\psi)}
d\bar{\psi} d\psi
\label{5}
\ee
is not gauge invariant
\be
\frac{\det \hat{\nabla}(A^g)}{\det \hat{\nabla}(A)}
=\hbox{e}^{{i}\a_1 (A,g)},
\label{6}
\ee
and
\bea
\alpha_1 (A,g)&=&\int d^4x\,[d^{-1}\kappa (g)-\frac {i}{48\pi^2}
\epsilon^{\mu \nu \lambda \sigma }\tr[(\am \partial_\nu \al
\,+\,\partial_\mu \an \al \,+\,\am \an \al )g_\sigma \,-\nonumber \\
&&-\,\frac 12\am g_\nu \al g_\sigma \,-\,\am g_\nu g_\lambda
g_\sigma]]
\label{7}
\eea
we use the notations
\be
\int d^4xd^{-1}\kappa (g) \equiv -\frac{i}{240\pi^2}\int_{M_5} d^5x \,
\epsilon^{pqrst} \tr{(g_p
g_q g_r g_s g_t)}
\label{8}
\ee
\be
g_\mu=\partial_\mu gg^{-1}.
\label{9}
\ee
In eq.(\ref{8}) the  integration goes  over a  five-dimensional manifold
whose boundary is the usual  four-dimensional space.

The particular form of local ($mod\, 2\pi $) functional $\a_1 (A,g)$
depends essentially on the computation scheme (regularization) used for
calculation of the determinant but it cannot be annihilated in
any admissible scheme. In eq. (\ref{5}) a special choice of computation
scheme is used. This choice spoils the whole $SU(N)$ gauge symmetry
opposite  to one of ref.\cite{12} where a maximal subgroup isomorphic to
$SO(N)$ is preserved.

 From eq.(\ref{6}) one can easely see that $\a_1 (A,g)$ satisfies ($mod\,
2\pi $) identity (1-cocycle condition):
\be
\a_1 (A^h,g)-\a_1 (A,hg)+\a_1 (A,h) = 0, \quad h,g \in SU(N),
\label{10}
\ee
and difference given by a different computational scheme consists in
adding to $\a_1$ the gauge variation of a local (finite)
counterterm (trivial 1-cocycle) -- \be \a_0(A^g)-\a_0(A).  \ee

As a consequence of the gauge non-invariance of the determinant (\ref{5})
the modification of the constraint algebra (\ref{2},\ref{3}) occurs. In
particular the commutator for quantum operators $G^a$ will acquire
Schwinger term (infinitesimal 2-cocycle) $a^{ab}$ \be [G^a({\bf
x}),G^b({\bf y})]=if^{abc} \d ({\bf x-y}) + a^{ab}({\bf x,y}) \label{11}
\ee
which generally does not vanish on the constraint surface $G=0$. This fact
makes condition (\ref{4}) inconsistent and in this case one is stressed to
loose either Lorentz (and gauge) invariance or unitarity.

To repair this situation and have a gauge invariant quantum theory one can
modify the quantization procedure as proposed by L.D.Faddeev and
S.L.Shatashvili \cite{9}. According to this one must consider a modified
action
\be
S_{mod}=S_{cl}+\a_1 (A,g)
\label{12}
\ee
instead of the classical one and quantize it after imposing gauge condition
($A_0=0$). If the new action (\ref{12}) is well defined i.e. it has
nondegenerate symplectic form one has restored the gauge invariance of the
quantum theory with constraints $G$ obeyng the old algebra (\ref{2},\ref{3}).
Also one has a number of additional degrees of freedom carried by the
gauge group valued fields $g$ and the path integral representation for the
generating functional is given by
\be
Z=\int dA d\phi \d (A_0) [\det{\omega (A^g)}]^{1/2}\hbox{e}^{iS+\a_1(A,g)}.
\label{nondeg}
\ee
where $\det \omega (A^g) $ is the determinant of the symplectic form
\cite{9},
\be
\omega (A^g)= -\frac{i}{48 \pi^2} \e_{ijk} \tr \{ \d gg^{-1} \d
gg^{-1}(A^g_i\p_jA^g_k+\p_iA^g_jA^g_k+A^g_iA^g_jA^g_k)- \d gg^{-1} \p_iA^g_j
\d gg^{-1}A^g_k
\}.  \label{ndsf} \ee

But as was mentioned in the Introduction for some $SU(N)$ groups this
action can have degenerate symplectic form. This means that there are a
number of additional primary constraints that can generate for example
some secondary constraints and so on \cite{fss2,fss3}.
In this case one should modify the Faddeev-Shatashvili method to include
the whole tower of the constraints. In particular one must verify if the
additional constraints do not destroy the gauge invariance of the quantum
theory.

\section{Degenerate case}
To quantize the action (\ref{12}) with degenerate symplectic form let us
firstly introduce a parametrization of the gauge group element $g \in G
\equiv SU(N)$ by fields $\phi^A,\, A=1, \dots ,dim\, G$
\be
g=g(\phi)
\label{13}
\ee
In terms of these fields the action of the theory in first order formalism
looks as follows
\bea
S&=& \int d^4 x \{ E_i^a\dot{A}_i^a -\frac{i}{48 \pi^2} \e_{ijk} \tr
\dot{A}_i\{A_j ,g_k\}+ \Gamma_A\dot{\phi}^A \nonumber \\
&+& \frac{i}{48\pi^2}\e_{ijk} \tr (A_i\p_j A_k + \p_iA_jA_k +A_iA_jA_k +
A_ig_jA_k -A_ig_jg_k)g_A \dot{\phi}^A \nonumber \\
&+& i \psi^{+} \p_0 \psi - \frac12(E^2+B^2)- i\bar{\psi} \gamma_i \nabla_i
\psi + A_0^aG^a
\label{14}
\eea
where $E$ and $B$ are "electric" and "magnetic"  components of the gauge
field strenght $F_{\mu \nu}$, $\Gamma^A \dot{\phi_A}$ is the term of
eq.(\ref{8}), $g_A \equiv \frac{\d g(\phi)}{\d \phi^A}g^{-1}$, $\e_{ijk}$
is three dimensional antisymmetric tensor ($i=1,2,3$). From eq.(\ref{14})
one can see that one has a set of constraints -- Gauss law \bea
G^a&=&\nabla_iE_i+i\psi^+\lambda^a \psi-\frac{i}{48 \pi^2} \e_{ijk} \tr
\lambda^a (2\{ \p_iA_j,g_k\}-g_ig_jA_k - \nonumber \\ &-&A_ig_jg_k
+A_iA_jg_k-A_ig_jA_k-g_iA_jA_k-g_iA_jg_k-g_ig_jg_k \label{17} \eea As we
have already mentioned we impose the gauge condition $A_0=0$ this will
exclude the constraints $G^a$ from our analysis on the classical level.

Canonical momenta for the fields $A^a_i $ are given by shifting of $E$
\be
E^a_i \rightarrow \Pi^a_i =E^a_i- \frac{i}{48 \pi^2} \e_{ijk} \tr \lambda^a
\{A_j,g_k\}
\label{18}
\ee
Introducing  also the canonical momenta for the fields $\phi^A$ one gets a
set of constraits
\bea
\varphi_A&=&p_A-\Gamma_A+ \nonumber \\
&+& \frac{i}{48\pi^2}\e_{ijk} \tr (A_i\p_j A_k + \p_iA_jA_k +A_iA_jA_k +
A_ig_jA_k -A_ig_jg_k)g_A
\label{19}
\eea
Also one has an equivalent set of constraints $\varphi_a$ given by
\be
\varphi_a=g_a^A\varphi_A=0
\label{20}
\ee
where $g_a^A$ is the inverse to the matrix $g_A^a=\frac12\tr
g_A\lambda^a$. The matrix of Poisson brackets of the constraints
(\ref{19}) is just the symplectic form for fields $\phi^A$
\bea
\omega_{AB}=\{\varphi_A({\bf x}), \varphi_B({\bf y})\} &=&
\frac{i}{48\pi^2}\e_{ijk} \tr (
\frac12[g_A,g_B](\wa_i\p_j\wa_k+\p_i\wa_j\wa_k+\wa_i\wa_j\wa_k) +
\nonumber \\ &+& \frac12(g_A\wa_ig_B\p_j\wa_k-g_B\wa_ig_A\p_j\wa_k)
) \d
({\bf x-y})
\label{21}
\eea
where $\wa_i \equiv (A_i+g_i)$.  But it will
be more convenient to use the equivalent set of constraints (\ref{20}) for
which the matrix of Poisson brackets is --
\bea
\omega_{ab}=\{\varphi_a({\bf x}),
\varphi_b({\bf y})\} &=& \frac{i}{48\pi^2}\e_{ijk} \tr (
\frac12[\lambda_a,\lambda_b](\wa_i\p_j\wa_k+\p_i\wa_j\wa_k+\wa_i\wa_j\wa_k) +
\nonumber \\ &+&
\frac12(\lambda_a\wa_i\lambda_b\p_j\wa_k-\lambda_b\wa_i\lambda_a\p_j\wa_k)
) \d ({\bf x-y})
\label{22}
\eea

To write the transformation law of (\ref{22}) one can observe that
$g\lambda_a g^{-1}$ can be expanded in terms of $\lambda_a$
\be
\lambda_a \rightarrow g\lambda_ag^{-1}={\Lambda_a}^b (g)\lambda_b, \quad
\det \Lambda =1
\label{23}
\ee
 From eqs.(\ref{22},\ref{23}) one can see
that the matrix (\ref{22}) is transformed under gauge group action as
follows
\be
\omega_{ab} ({\bf x})\rightarrow {\Lambda_a}^c({\bf x})
\omega_{cd}({\bf x}){\Lambda_b}^d({\bf x}),
\label{24}
\ee where $\omega_{ab} ({\bf x,y})=\omega_{ab} ({\bf x}) \d({\bf x-y})$.

If the matrix $\omega_{ab}$ is a degenerate one it has
a set of linearly independent null vectors $z^b_R$, $R=1,...,K<N \equiv
dimG$
\be
\omega_{ab}z^b_R(\wa)=0 ,
\label{25}
\ee
where $K$ is corank of the matrix $\omega_{ab}$.

One can find these null
vectors as follows. Consider the following antisymmetric (isotopic) tensor
\be
z^{a_1\dots a_K}({\bf x})=\e^{a_1\dots
a_Kb_1b_2\dots b_{N-K-1}b_{N-K}} \omega_{b_1 b_2}({\bf x})\dots
\omega_{b_{N-K-1}b_{N-K}}({\bf x}) \label{ns} \ee

This tensor has the following properties
\be
z^{a_1\dots a_K}({\bf x})\omega_{a_1 b}=0
\label{prop1}
\ee
\be
z^{aa_2\dots a_K}({\bf x}) z_{ba_2\dots a_K}({\bf x}) \sim  {P^a}_b
\label{prop2}
\ee
where ${P^a}_b$ in the last equation stands for the projector on zero
subspace of the matrix $\omega $.
To prove (\ref{prop1}) one can consider a maximal nonzero minor
$\omega_{\bar{a} \bar{b}}$. All other lines and columns of the matrix
$\omega_{ab}$ are linear combinations of elements $\omega_{\bar{a}
\bar{b}}$. One can see  that only they contribue for eq.(\ref{ns}). In
fact $z^{a_1\dots a_K}$ is skew product of all $\omega_{\bar{a} \bar{b}}$.
So contracting $z$ with $\omega_{ab}$ one will have antisymmetric
combination which is equal to zero because there is a component
$\omega_{\bar{a}
\bar{b}}$ which meets itself at least twice in this antisymmetric product.
Another way to prove eq. (\ref{prop1}) is to use Darboux theorem.
Eq. (\ref{prop2}) can be obtained by expansion of product of two $\e $ in
antisymmetric combinations of delta symbols.

It is clear that for any set of null
vectors $\{z_R\}$ the skew product of these vectors must be proportional
to the tensor (\ref{ns})
\be
\e^{P\dots R} z^{a_1}_P\dots z^{a_K}_R \sim z^{a_1\dots a_K}
\label{iden}
\ee
where $\e^{P\dots R}$ is $K$ dimensional antisymmetric symbol with
$\e^{1\dots K}=1$, and viceversa any set of linear independent vectors
satisfying (\ref{iden}) is set of null vectors of $\omega $.

In particular if there is the only null vector it is given by
\be
z^{a}({\bf x})=\e^{ab_1b_2\dots b_{N-2}b_{N-1}}
\omega_{b_1 b_2}({\bf x})\dots \omega_{b_{N-2}b_{N-1}}({\bf x})
\label{nv}
\ee
In this case the gauge group must be odd dimensional.

In  case when there are more than one constraint one has the
combinations of constraints
\be
\psi^{a_2\dots a_K}({\bf x})= z^{a_1a_2\dots a_K}({\bf x})
\varphi_{a_1}({\bf x})
\label{cc}
\ee
which commute with all the constraints $\varphi_a$ but they are not
all independent.

So one must either introduce a set of linear independent vectors
satisfying (\ref{iden}) or equivalently select the subset of constraints
from (\ref{cc}).

Now suppose that one has choosen such set of null vectors $z^a_R$ as
solution of the equation
\be
\e^{P\dots R} z^{a_1}_P\dots z^{a_K}_R = z^{a_1\dots a_K}.
\label{qq}
\ee
These vectors are invariant under gauge group action
\be
z^a_R \rightarrow  {(\Lambda^{-1})^a}_bz^b_R \label{26}
\ee
So the set of independent commuting constraints consist of
\be
\varphi_Q=z^a_Q\varphi_a.
\label{prcon}
\ee

Now following the Dirac procedure one should impose the conservation of
these constraints getting in this way the secondary constraints,
\be
\tf_Q({\bf x}) \equiv \{H,\varphi_Q({\bf x})\}= z_Q^a({\bf
x})\{H,\varphi_a({\bf x})\}=0,
\label{seccon}
\ee
where equalities are hold up
to the primary constraint combinations and
\bea
&&\{H,\varphi_a\}=\nonumber \\
&&-\frac{i}{48\pi^2}\e_{ijk} \tr
\lambda^a(\{E_i,F_{jk}\}-E_i\wa_j\wa_k+\wa_iE_j\wa_k -\wa_i\wa_jE_k).
\label{hcon}
\eea
One can see that the commutator (\ref{hcon}) transforms covariantly under
gauge group action
\be
\{H,\varphi_a\} \rightarrow {\Lambda_a}^b \{H,\varphi_b\}.
\label{transf}
\ee
Due to this the secondary constraints will be invariant under gauge group
transformations.
This shows that imposing of the
secondary constraints will not destroy the gauge invariance if the
determinant of matrix of all constraints Poisson brackets is also
gauge invariant.

Now one can unify the secondary constraints with the primary ones to get
the set of constraints $\varphi_I$, $(I)=(a,Q)$ and consider the matrix of
the Poisson brackets of this set of constraints. It has the following block
structure
\be
||\omega_{IJ}||=\left( \begin{array}{cc}
\{\varphi_a,\varphi_b\} & \{\varphi_a,\tf_Q\} \\
-\{\varphi_a,\tf_P\} & \{\tf_Q,\tf_P\} \end{array} \right)
\label{newsf}
\ee

Let us prove firstly the gauge invariance of its determinant. In order to
do this consider its gauge transformation law.

The transformation law of the block $ \{\varphi_a,\varphi_b\} $ is given
by (\ref{24}). To find the transformation law of the block
$\{\varphi_a,\tf_Q\}$ let us remind that the constraints have the
following structure
\be
\tf_Q=\tr E_i\Phi_{iQ}=\tr E_i\Phi_{ib}z^b_Q
\label{phiq}
\ee
where
\be
\Phi^b_{ia}({\bf x})\d({\bf x-y})=\{E^b_i({\bf x}),\varphi_a({\bf y})\}
\label{phiq2}
\ee
is gauge covariant.

So the Poisson bracket has the form
\be
\{\varphi_a,\tf_Q\}=\{\phi_a,E^a_i\}\Phi^a_{iQ}+E^a_ig^A_a\{p_A,\Phi^a_{iQ}\}
\label{block}
\ee
The first term in (\ref{block}) is equal to $\Phi^a_{iQ}\Phi^a_{ib}\d
({\bf x-y})$ and it is gauge covariant (in indice $a$). The remaining term
can be rewritten as follows
\be
E^a_i g^A_a\{p_A,\Phi^b_{iQ}(\wa )\}=E^a_k g^A_a\frac{\d \Phi^b_{kQ}(\wa
)}{\d \wa_i } \frac{\d g_i}{\d \phi^A},
\label{block1}
\ee
where
\be
g^A_a \frac{\d g_i}{\d \phi^A}=\lambda_a \d^\prime_j({\bf x-y})+g^B_a
\nabla_i^gg_B \d({\bf x-y})
\label{bl2}
\ee
\be
\nabla_i^gg_B  \equiv \p_ig_B-[g_i,g_B]
\label{bl3}
\ee
Since (\ref{bl2}) transforms covariantly it follows that the last term in
(\ref{block}) is also covariant.

Unfortunately proving the gauge invariance of the last block $ \{
\tf_P,\tf_Q\}$ is not so easy. It probably holds only on
the surface of all second class constraints but luckily the
determinant of $\omega $ does not depend on this block.

Indeed let us consider the following antisymmetric block matrix which can
be obtained from our by splitting the set of constraints $\varphi_a$ in
the subset of commuting constraints $\varphi_Q$ and the subset of second
class ones $\varphi_{\bar{a}}$
\be
\left( \begin{array}{ccc} \{\varphi_{\bar{a}} ,\varphi_{\bar{b}}\}
 & \{\varphi_{\bar{a}},\varphi_S \}& \{\varphi_{\bar{a}},\tf_Q\} \\
\{\varphi_R,\varphi_{\bar{b}}\} & \{\varphi_R, \varphi_S\}
& \{\varphi_R,\tf_Q\} \\
\{\tf_P, \varphi_{\bar{b}}\}&\{\tf_P, \varphi_S\} & \{\tf_P,\tf_Q\}
\end{array} \right) \equiv
\left( \begin{array}{ccc} \widetilde{\omega} & 0 & a
\\ 0 & 0 & b \\ -a^T & -b^T & c \end{array} \right)
\label{det}
\ee with
nondegenerate block $\widetilde{\omega}=-\widetilde{\omega}^T$, $c=-c^T$
and quadratic block $b$. One can see that this matrix is
degenerate if and only if the block $b$ is degenerate. Using the
properties of the determinant one has \be \det \left( \begin{array}{ccc}
\widetilde{\omega} & 0 & a \\ 0 & 0 & b \\ -a^T & -b^T & c \end{array}
\right) = \det \left( \begin{array}{ccc} \widetilde{\omega} & 0 & 0 \\ 0 &
0 & b \\ 0 & -b^T & 0 \end{array} \right) =\det \widetilde{\omega} (\det
b)^2 \label{det2} \ee i.e. it does not depend on the block $c$.

Since the relevant part of $\omega$ transforms by multiplicaton on
orthogonal matrix the gauge invariance of the determinant is proved.

Let us note that Poisson brackets of the primary constraints are
ultralocal i.e. they contains only delta functions and not its
derivatives. From the other hand the Poisson brackets containing
secondary constraints are not ultralocal. They contains not only "pure"
delta functions but the derivatives of the delta function
are also present.
This feature could complicate the further analysis in case when $\omega$
is degenerate. Indeed, to find now the null vectors of such matrix one
have to solve a system of differetial equations. And the absence of
the manifest gauge invariance of $\omega$ itself will still complicate
the proving of the gauge invariance (if it exists) of the final theory.

Fortunately there are some indications that in a general position point
 the procedure of the reproduction of the constraints stop here.

We know that $\omega$ is degenerate
if and only if the block
 $b_{PQ}=\{\varphi_P,\tf_Q\}$ is degenerate.  Now consider the subspace
 of phase space such that $E=0$.  Note that this subspace belongs to the
 surface of the secondary constraints. So consider the matrix $b$ on this
 subspace. There it has the following form (see eq.(\ref{block})) \be
 b_{PQ}=\tr \Phi_{iP} \Phi_{iQ} \label{e=0} \ee and if constraints
 $\tf_Q=\tr E_i \Phi_{iQ}$ are independent in the vicinity of  some
point of phase space which is natural then (\ref{e=0})must be
nondegenerate in this point. Moreover this will hold nearly everywhere
due to the fact that determinant is a polinomial in the fields.

Considering $\omega$ to be nondedegenerate one can write the path integral
representation for the generating functional
\be
Z=\int \exp i(\Pi_i
\dot{A}_i+p_A\dot{\phi}^A-\frac12(E^2+B^2)+S_f(\bar{\psi},\psi,A))\d(\varphi_I)
\det||\omega_{IJ}({\bf x-y}) ||d\Pi dAdp d\phi d\bar{\psi} d\psi
\ee
where $S_f$ is the fermionic part of the action. Inegrating over $p$ and
changing variables back  $\Pi \rightarrow  E$ \be Z=\int \exp
i(S(\bar{\psi},\psi,A)+\a_1(A,g))\d(\varphi_Q) \det||\omega_{IJ}({\bf
x-y}) ||dEdA d\phi d\bar{\psi} d\psi \label{z} \ee Since gauge invariance
is restored one can choose the physical subspace by imposing (\ref{5}) \be
G^a|ph>=0
\ee
with physical observables having the form
\be
F(\Phi,g)=F(\Phi^h,h^{-1}g)=F(\Phi^g)
\label{obs}
\ee
where $\Phi = (A,E,\bar{\psi},\psi)$.

The number of the Lagrangean degrees of freedom can be established in the
 following way. It is equal to the number of (Lagrangean) fields minus
 half number of the second class constraints. In the case when only the
 secondary constraints appear one has
 \be
 N_{df}=N_A+(N_p+N_{phi})-\frac12 N_{\varphi}+N_{fermions}=
 2N+\frac12(N-K)+N_{fermions}
 \ee
 where $N=dim G $ and fermionic degrees of freedom are denoted as
$N_{fermions}$.

{\bf Discutions:} We have shown that despite of degeneracy of the
symplectic form of Wess-Zumino action one can perform the canonical
quantization of the anomalous  $SU(N)$ Yang-Mills theory and obtain
finally a gauge invariant quantum theory with new degrees of freedom and
additional constraints caused by degeneracy.

One can alternatively consider the initial anomalous theory with the
constraints $G_a$ satisfying algebra (\ref{11}) and canonically quantize
it. In our approach it corresponds to the quantization in $g=1$ gauge.
Due to the coincidence
\be
\omega_{ab}|_{g=1}=a_{ab}
\label{coin1}
\ee
one has for $a$ the same set of null vectors $z_Q(A)$. One can calculate
the commutator $\{H, G^a\}$, for example, using Bjorken-Johnson-Low
formula in a manner similar to calculation of the commutator $\{G,G\}$ in
ref.\cite{amfs}. Performing this one will have for it
\be
\{H,G_a\}=-\frac{i}{48\pi^2}\e_{ijk} \tr
\lambda^a(\{E_i,F_{jk}\}-E_iA_jA_k+A_iE_jA_k
-A_iA_jE_k)=\{H,\varphi_a\}|_{g=1}.
\ee
After this one can write the path integral representation for generating
functional
\be
Z=\int d\Phi \hbox{e}^{iS} \d (G_a) \det a
\label{g=1}
\ee
Using Faddeev-Popov trick one can formally pass from this integral to
one in the gauge $A_0=0$ obtaining the result identic to eq.(\ref{z}).
The last quantization procedure will always give the gauge invariant theory
because the shifting of the fields by gauge transformation is performed
only after the quantization. But to write the Poisson brackets one uses
formulas from quantum commutators which is rather formal procedure.
Nevertheless in the case when our first sheme gives the gauge invariant
theory last one must agree with it.

{\bf Aknowledgements:} I would like to thank A.A.Slavnov and
S.A.Frolov for useful discussions. This work was partially supported by
 RBRF grant under No.94-01-00300a and ISF grant under No.MNB000.

\end{document}